\documentclass[twoside,twocolumn,9pt]{article}
\usepackage{extsizes}
\usepackage[super,sort&compress,comma]{natbib}
\usepackage[version=3]{mhchem}
\usepackage[left=1.5cm, right=1.5cm, top=1.785cm, bottom=2.0cm]{geometry}
\usepackage{balance}
\usepackage{times,mathptmx}
\usepackage{sectsty}
\usepackage{graphicx}
\usepackage{lastpage}
\usepackage[format=plain,justification=justified,singlelinecheck=false,font={stretch=1.125,small,sf},labelfont=bf,labelsep=space]{caption}
\usepackage{float}
\usepackage{fancyhdr}
\usepackage{fnpos}
\usepackage[english]{babel}
\usepackage{array}
\usepackage{droidsans}
\usepackage{charter}
\usepackage[T1]{fontenc}
\usepackage[usenames,dvipsnames]{xcolor}
\usepackage{setspace}
\usepackage[compact]{titlesec}

\definecolor{cream}{RGB}{222,217,201}

\usepackage{amsmath}
\usepackage[version=3]{mhchem}
\usepackage{braket}
\def\bq{{\bf q}}
\def\bx{{\bf x}}
\def\id{{\textrm d}}

\begin{document}

\pagestyle{fancy}
\thispagestyle{plain}
\fancypagestyle{plain}{

\fancyhead[C]{\includegraphics[width=18.5cm]{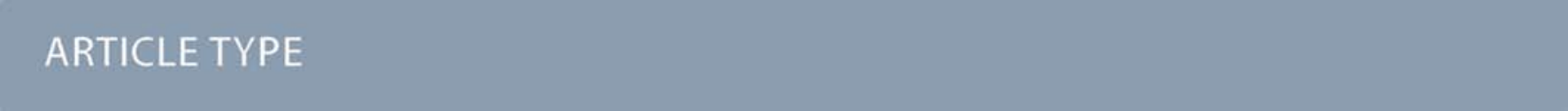}}
\fancyhead[L]{\hspace{0cm}\vspace{1.5cm}\includegraphics[height=30pt]{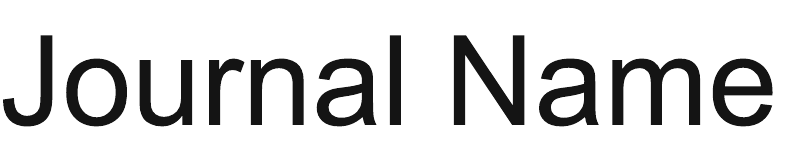}}
\fancyhead[R]{\hspace{0cm}\vspace{1.7cm}\includegraphics[height=55pt]{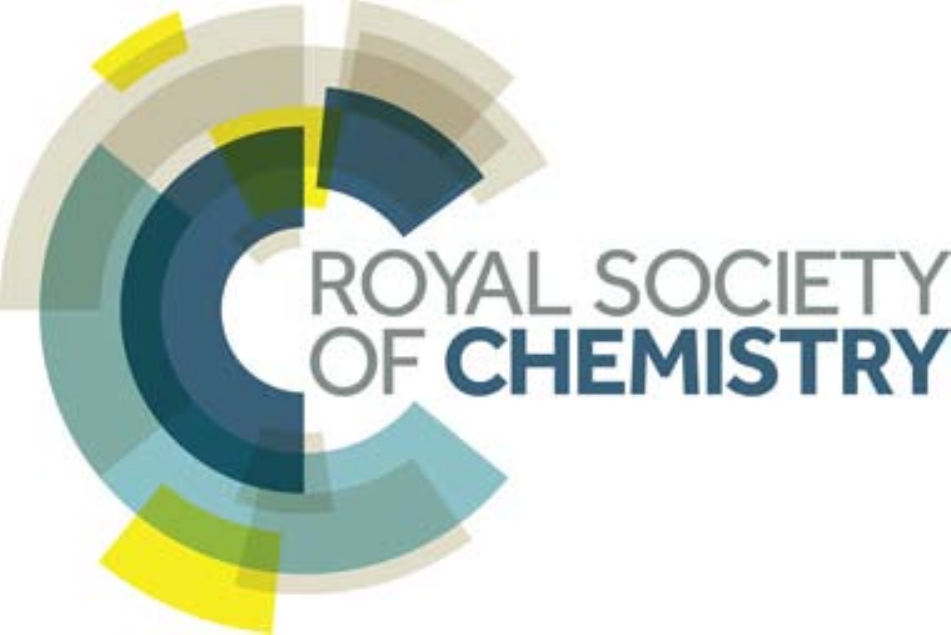}}
\renewcommand{\headrulewidth}{0pt}
}

\makeFNbottom
\makeatletter
\renewcommand\LARGE{\@setfontsize\LARGE{15pt}{17}}
\renewcommand\Large{\@setfontsize\Large{12pt}{14}}
\renewcommand\large{\@setfontsize\large{10pt}{12}}
\renewcommand\footnotesize{\@setfontsize\footnotesize{7pt}{10}}
\makeatother

\renewcommand{\thefootnote}{\fnsymbol{footnote}}
\renewcommand\footnoterule{\vspace*{1pt}%
\color{cream}\hrule width 3.5in height 0.4pt \color{black}\vspace*{5pt}} 
\setcounter{secnumdepth}{5}

\makeatletter 
\renewcommand\@biblabel[1]{#1}            
\renewcommand\@makefntext[1]%
{\noindent\makebox[0pt][r]{\@thefnmark\,}#1}
\makeatother 
\renewcommand{\figurename}{\small{Fig.}~}
\sectionfont{\sffamily\Large}
\subsectionfont{\normalsize}
\subsubsectionfont{\bf}
\setstretch{1.125} 
\setlength{\skip\footins}{0.8cm}
\setlength{\footnotesep}{0.25cm}
\setlength{\jot}{10pt}
\titlespacing*{\section}{0pt}{4pt}{4pt}
\titlespacing*{\subsection}{0pt}{15pt}{1pt}

\fancyfoot{}
\fancyfoot[LO,RE]{\vspace{-7.1pt}\includegraphics[height=9pt]{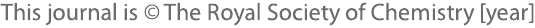}}
\fancyfoot[CO]{\vspace{-7.1pt}\hspace{13.2cm}\includegraphics{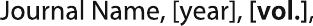}}
\fancyfoot[CE]{\vspace{-7.2pt}\hspace{-14.2cm}\includegraphics{head_foot/RF}}
\fancyfoot[RO]{\footnotesize{\sffamily{1--\pageref{LastPage} ~\textbar  \hspace{2pt}\thepage}}}
\fancyfoot[LE]{\footnotesize{\sffamily{\thepage~\textbar\hspace{3.45cm} 1--\pageref{LastPage}}}}
\fancyhead{}
\renewcommand{\headrulewidth}{0pt} 
\renewcommand{\footrulewidth}{0pt}
\setlength{\arrayrulewidth}{1pt}
\setlength{\columnsep}{6.5mm}
\setlength\bibsep{1pt}

\makeatletter 
\newlength{\figrulesep} 
\setlength{\figrulesep}{0.5\textfloatsep} 

\newcommand{\topfigrule}{\vspace*{-1pt}%
\noindent{\color{cream}\rule[-\figrulesep]{\columnwidth}{1.5pt}} }

\newcommand{\botfigrule}{\vspace*{-2pt}%
\noindent{\color{cream}\rule[\figrulesep]{\columnwidth}{1.5pt}} }

\newcommand{\dblfigrule}{\vspace*{-1pt}%
\noindent{\color{cream}\rule[-\figrulesep]{\textwidth}{1.5pt}} }

\makeatother

\twocolumn[
  \begin{@twocolumnfalse}
\vspace{3cm}
\sffamily
\begin{tabular}{m{4.5cm} p{13.5cm} }

\includegraphics{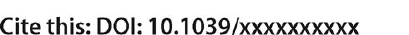} & \noindent\LARGE{\textbf{Unraveling the dominant phonon scattering mechanism in thermoelectric compound ZrNiSn}} \\
\vspace{0.3cm} & \vspace{0.3cm} \\

 & \noindent\large{Ankita Katre, Jes\'{u}s Carrete and Natalio Mingo} \\

\includegraphics{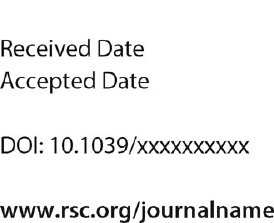} & \noindent\normalsize{
Determining defect types and concentrations remains a big challenge of semiconductor materials science. 
By using \emph{ab-initio} thermal conductivity calculations we reveal that Ni/vacancy antisites, 
and not the previously claimed Sn/Zr antisites, are the dominant defects affecting thermal transport in half-Heusler compound 
ZrNiSn. Our calculations correctly predict the thermal conductivity dependence with temperature and 
concentration, in quantitative agreement with published experimental results. Furthermore, we find a 
characteristic proportionality between phonon-antisite scattering rates and the sixth power of phonon 
frequency, for which we provide an analytic derivation. These results suggest that thermal conductivity 
measurements in combination with \emph{ab-initio} calculations can be used to quantitatively assess 
defect types and concentrations in semiconductors.}
\end{tabular}

 \end{@twocolumnfalse} \vspace{0.6cm}

  ]

\renewcommand*\rmdefault{bch}\normalfont\upshape
\rmfamily
\section*{}
\vspace{-1cm}


\footnotetext{\textit{LITEN, CEA-Grenoble, 17 rue des Martyrs, 38054 Grenoble Cedex 9, France; E-mail: ankitamkatre@gmail.com, jcarrete@gmail.com, natalio.mingo@cea.fr}}


\section{\label{sec:intro} Introduction}

Defects are ubiquitous in a material. In recent years, the role of defects on the mechanical, electronic and 
thermal properties of semiconductor materials has being intensely studied, particularly regarding energy 
applications.\cite{Freysoldt_2014,Hu_JMCA2015, Zandi_NComm2014, Doak_JMCC2015, Xu_JMCA2013, Cao_JMCA2015, Walle_JAP2004, Liu_JMCA2015}
Half-Heusler compounds have attracted a lot of interest for applications as 
thermoelectrics and solar cells.\cite{Casper_SST2012, Chen_MT2013} 
Improvements in their thermoelectric figure of merit $zT=\frac{S^2\sigma}{\kappa}T$ -- where T represents 
temperature, $S$ the Seebeck coefficient,
$\sigma$ the electrical conductivity and $\kappa$ the thermal conductivity comprising both the lattice
($\kappa_{\ell}$) and electronic ($\kappa_\text{e}$) contributions -- have been achieved mostly for
multicomponent half-Heusler alloys with complex microstructures.\cite{Fecher_PSSA2016,
Sakurada_APL2005, Shen_APL2001, Ouardi_APL2010, Krez_JMCA2014, Sootsman_AC2009, Yu_AM2009, Qiu_APL2010, Xie_CEC2012} 

\begin{figure}[t]
 \centering
  \includegraphics[width=8.3cm]{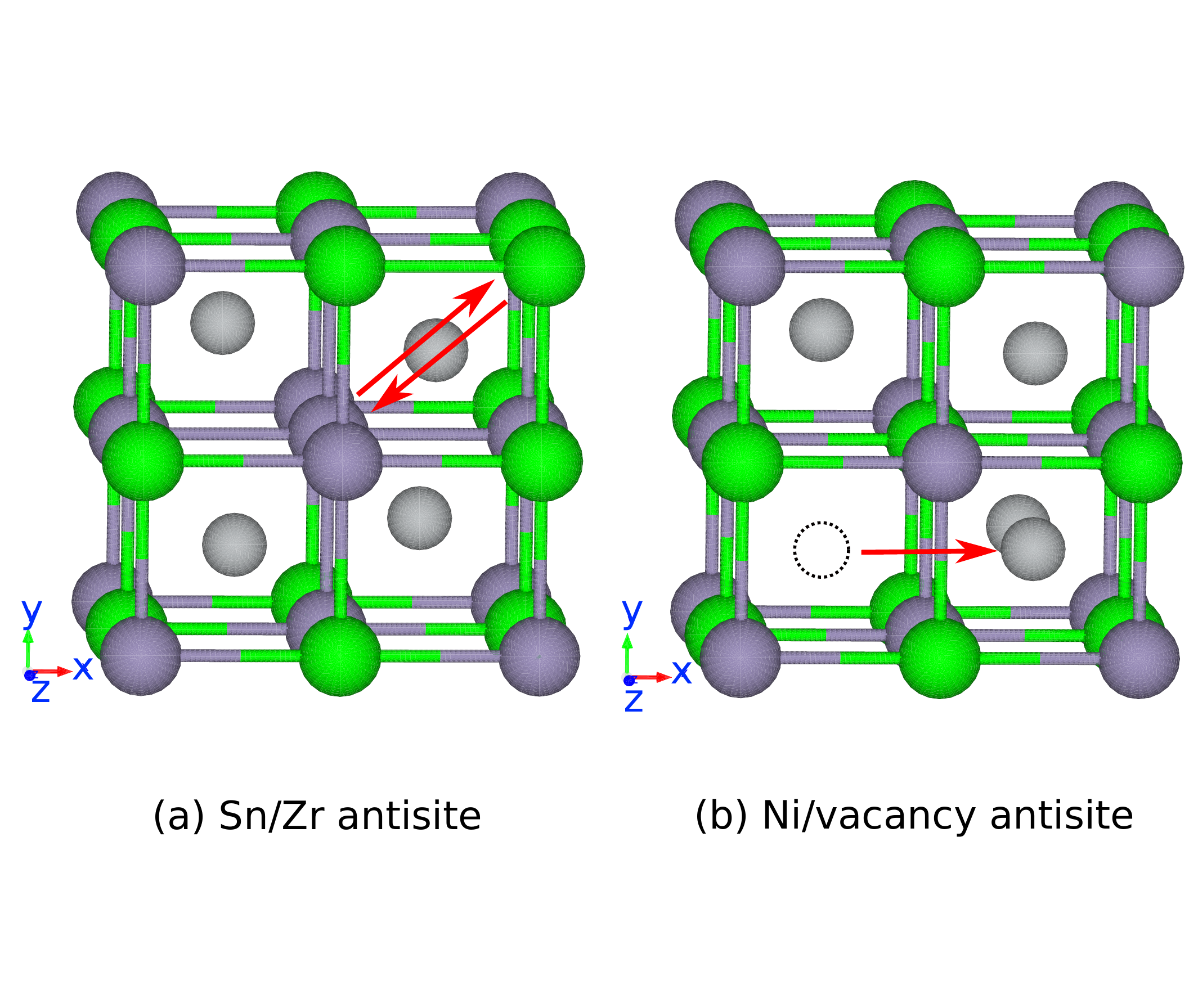}
  \caption{ Representation of (a) Sn/Zr antisites and (b) Ni/vacancy in ZrNiSn (Zr: Green atoms, Sn: 
  purple atoms, Ni: small grey atoms).}
  \label{fig:antisite_reffig}
\end{figure}

Despite the tremendous importance of defects on materials' technological applications, determining the 
concentration of different defects and their effects is still a very challenging problem.\cite{Freysoldt_2014} 
One such example is of the thermoelectric half-Heusler ZrNiSn. In this material, 
$\kappa$ decreases in the presence of antisite defects, but different conflicting  
experimental results are inconclusive about which antisite has dominant effect on 
the $\kappa$.\cite{Qiu_APL2010, Xie_CEC2012} 
There are two kinds of antisites in ZrNiSn which occur inherently during synthesis namely Sn/Zr and Ni/vacancy antisites.
Due to similar atomic radii of Sn ($1.58$ \AA) and Zr ($1.60$ \AA), 
Sn/Zr antisites can be present in ZrNiSn (Fig.~\ref{fig:antisite_reffig}a).\cite{Aliev_ZPB1989} 
In addition, Ni can migrate to one of the four vacant pockets (out of a total of eight) 
in the Sn-Zr rocksalt structure which can be considered as 
generalized Ni/vacancy antisite (Fig.~\ref{fig:antisite_reffig}b).\cite{Qiu_APL2010, Xie_CEC2012}

In this work, with the aid of \emph{ab-initio} calculations, we reveal the dominant phonon scattering behaviour of 
Ni/vacancy over Sn/Zr antisites, contradicting the claim of an earlier study by Qiu \emph{et al.}\cite{Qiu_APL2010} 
They measured $\kappa$ for unannealed and annealed samples of ZrNiSn prepared by arc-melting
method. They found that the $\kappa$ at $300\,\mathrm{K}$ increases
from $\sim 7\,\mathrm{W/m/K}$, for the unannealed sample, to almost twice that value after one week of
annealing.\cite{Qiu_APL2010} They attributed this result to a strong decrease of Sn/Zr antisite concentration in their annealed sample  
as compared to the unannealed one. However, no mention of the antisite concentration is found in their work. In contrast, a
similar $\kappa$ was measured by Xie \emph{et al.} for unannealed samples prepared by levitation
melting method, which they claimed contains no trace of Sn/Zr but $\sim 4.3\%$ at $300\,\mathrm{K}$ of Ni/vacancy
antisites.\cite{Xie_CEC2012} Furthermore, annealing for one week (down to a $2.9\%$ Ni/vacancy
concentration) increased $\kappa$ only by $\sim 15\%$. 
While these experiments
point to a reduction in $\kappa$ with increasing antisites, they also raise a number of
questions regarding the role of each of the antisites in this material, such as:
1. Which of the two antisite is the dominant phonon scatterer in ZrNiSn?
2. What are the scattering strengths of each of these antisites?
3. How much do they affect the thermal conductivity of ZrNiSn?

In this paper, we calculate $\kappa_{\ell}$ for ZrNiSn from first principles using 
predictive approach for antisite scattering and explain 
the underlying phonon scattering behaviour to answer the above mentioned questions. 

\section{\label{sec:mthd}Computational methodology}
For a cubic
system, the expression for $\kappa_{\ell}$, obtained by solving the linearized Boltzmann transport
equation in the relaxation time approximation, is given by an integral over phonon wave-vectors $\bf q$ 
in the Brillouin zone summed over phonon branches $j$, as \cite{Li_PRB2012, Katre_PRB2016}
\begin{equation}
 \kappa_{\ell} = \frac{1}{3}\sum_j \int \frac{\id\bq}{(2\pi)^3} C_{j\bq} v^2_{j\bq}\tau_{j\bq}
 \label{eq:RTAkappa}
\end{equation}
where $C_{j\bq}$ is the contribution to the heat capacity, $v_{j\bq}$ the group velocity and
$\tau_{j\bq}$ the relaxation time for phonon mode $j\bq$. Each phonon scattering rate, $1/\tau$ is
calculated as a sum of contributions from anharmonic processes and antisite defects,
\begin{equation}
 1/\tau_{j\bq} = 1/\tau^{\text{anh}}_{j\bq} + \sum_{\substack{\text{def = Sn/Zr,} \\ \text{Ni/vacancy}}} 1/\tau^{\text{def}}_{j\bq}.
 \label{eq:tau_tot}
\end{equation}
$1/\tau^{\text{anh}}_{j\bq}$ is obtained from first principles employing information about the second- and
third-order interatomic force constants (IFCs) in the compounds. Its expression can be found in 
Refs.~\citenum{Li_PRB2012, Katre_PRB2016}. Likewise, the expression for $1/\tau^{\text{def}}_{j\bq}$ comes from an
atomistic description of each defect, following the \emph{ab-initio} Green's function approach, as
explained in detail in Refs.~\citenum{Mingo_PRB2010, Katcho_PRB2014,Kundu_PRB2011}.
Using the Lippmann-Schwinger equation, $1/\tau^{\text{def}}_{j\bq}$ is given as,
\begin{equation}
  \begin{split}
    & 1/\tau^{\text{def}}_{j\bq} = f_{\text{def}} 
    \frac{\Omega}{V_{\text{def}}}\frac{1}{\omega_{j\bq}} \times \\
    & \sum_{j'\bq'} \left\vert \Braket{ j'\bq' | (\mathbf{I - \mathbf{\hat{V}}g^{+})^{-1}\mathbf{\hat{V}}} | j\bq } \right\vert^2
    \delta \left(\omega_{j'\bq'}^2-\omega_{j\bq}^2\right)
  \end{split}
  \label{eq:tau_def}
\end{equation}
where $\ket{j\bq}$ and $\ket{j'\bq'}$ represent the incident and outgoing phonons.  Here, $f_{\text{def}}$,
$V_{\text{def}}$ and $\Omega$ represent the concentration of defects, the volume of each defect, and the volume
in which $\ket{j\bq}$ is normalized, respectively, and $\omega$ is the angular frequency. The core part of
the equation contains the identity matrix $\bf{I}$, the retarded Green's function of the perfect crystal
$\bf{g^{+}}$, and the perturbation matrix $\bf{\hat{V}}$. The $\bf{\hat{V}}$ matrix is expanded as,
\begin{equation}
 \bf{\hat{V}} = \bf{\hat{V}_K} + \bf{\hat{V}_M}
\end{equation}
where $\bf{\hat{V}_K}$, $\bf{\hat{V}_M}$ are the contributions due to the change in the harmonic IFCs and
the masses respectively.\cite{Katcho_PRB2014}

For the calculation of the harmonic and anharmonic IFCs to obtain the
phonon frequencies and the scattering rates, we follow the finite displacement approach.  A
4$\times$4$\times$4 supercell based on the optimized primitive unit cell (space group F-43m,
$a = 5.99$\AA\ ) consisting of $192$ atoms is used for calculating the atomic forces. The force
calculations are performed using the projector-augmented-wave method\cite{paw} as implemented in the DFT
code VASP\cite{PAW_method} within the local density approximation and a $\Gamma$-only $k$-point mesh. The
Phonopy\cite{phonopy} package is used to extract the second order IFCs from the calculated
forces. Third-order IFCs are extracted using our own thirdorder.py code.\cite{Li_CPC2014} A non-analytic
correction to the dynamical matrix for reproducing the LO-TO phonon splitting in ZrNiSn is also included
by calculating the Born effective charges and the dielectric tensor with VASP.\cite{Wang_JPCM2010}
Green's functions are computed semi-analytically using the tetrahedron method for Brillouin-zone
integration.\cite{tetrahedron} The code implementing the tetrahedron method, as well as the calculation
of $\tau^{\text{anh}}_{j\bq}$, $\tau^{\text{def}}_{j\bq}$ from Eq.~\eqref{eq:tau_def} and $\kappa_{\ell}$, is
developed in house as a part of the ALMA project.\cite{ALMA_BTE} 

To calculate $1/\tau^{\text{def}}_{j\bq}$ for the case of an Sn/Zr antisite (mass ratio
$\frac{m_\text{Sn}}{m_\text{Zr}}\sim 1.3$, atomic radius ratio
$\frac{r_\text{Sn}}{r_\text{Zr}}\sim 1.01$) we use the virtual crystal approximation 
only considering the mass perturbation due to the exchange of atomic positions of Sn and Zr
(Fig.~\ref{fig:antisite_reffig}a) in one of the unit cells contained in the supercell. Thus in this case,
$\bf{\hat{V}} \simeq \bf{\hat{V}_M}$ is used in Eq.~\eqref{eq:tau_def}. Since there is no net change in mass
of the unit cell, but only an exchange of masses between two sites, the perturbation can be described as
a mass dipole. The applicability of virtual crystal approximation is also tested by relaxing the atomic
coordinates of the defective structure, which yields a maximum deviation of only $0.08$ \AA\ for
a few atoms around the defect.

The Ni/vacancy antisite is introduced in the 4x4x4 rhombohedral supercell by moving a Ni
atom by ($\frac{1}{2}, \frac{1}{2}, -\frac{1}{2}$) in terms of the primitive unit cell vectors, which
creates a defect parallel to the $OZ$ axis as shown in Fig.~\ref{fig:antisite_reffig}b. For this case,
$\bf{\hat{V}} = \bf{\hat{V}_K}$ is used in Eq.~\eqref{eq:tau_def}. We construct $\bf{\hat{V}_K}$ as the
difference between the second-order IFCs for the defective and pristine systems up to second nearest
neighbours for each atom within a cutoff of $8$ \AA\ (including $7$ neighbour shells) around the
defect. The cutoff is decided based on an analysis of the changes in atomic positions $> 0.01$~\AA\ after relaxing the
defect structure. The scattering rates are then calculated on a $28\times 28\times 28$ $q$-point mesh,
using a $18\times 18\times 18$ grid for the Green's function calculation.

\begin{figure}[t]
 \centering
  \includegraphics[width=8.3cm]{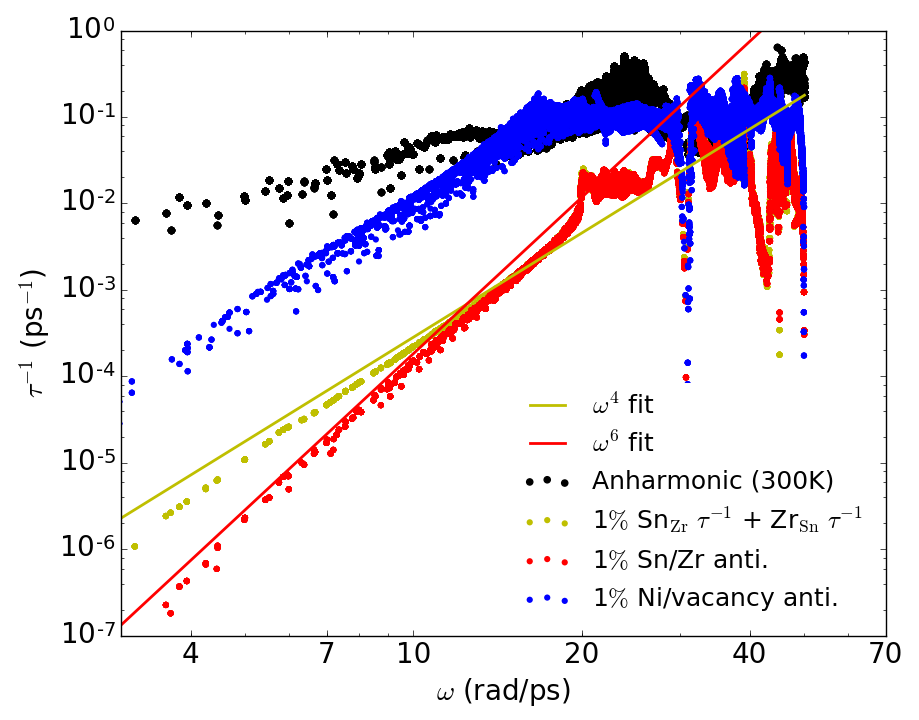}
  \caption{Phonon scattering rates by Sn/Zr antisite defects, by the sum of independent \ce{Sn_{Zr}} and \ce{Zr_{Sn}}
    point defects, by Ni/vacancy antisite defects and by anharmonic phonon interactions at
    $300\,\mathrm{K}$. Sn/Zr antisite defects are found to show an $\omega^6$ dependence, clearly different
    from the well know $\omega^4$ trend of point defects. Ni/vacancy antisites trend show slight 
    deviation from the exact $\omega^6$ dependence which could be attributed to the fact that 
    Ni/vacancy antisite is not a classic case of dipole defect.} 
  \label{fig:scat_rates}
\end{figure}

\section{\label{sec:res} Results and discussion}

\subsection{Antisite scattering rates in ZrNiSn.~~}

The calculated phonon-antisite scattering rates along with the anharmonic scattering at
$300\,\mathrm{K}$ for ZrNiSn are shown in Fig.~\ref{fig:scat_rates}. For comparison purposes, the figure
also shows the calculated Sn/Zr antisite scattering rate as the sum of the two
separate contributions from the mass perturbations resulting from the \ce{Sn_{Zr}} and \ce{Zr_{Sn}}
substitutions, describing a hypothetical independent antisite model.

At high frequencies the exact scattering rates are found to be very similar to the results of the
independent model, as seen in Fig.~\ref{fig:scat_rates}.  However, significant differences are
found at low frequencies where the common Rayleigh ($\omega^4$) behaviour of the scattering rates, that shows up in the results
of the independent antisite model, is replaced by an unconventional $\omega^6$ trend of the actual
results. As we show below, this unconventional dependence can be understood as the result of a dipole effect on phonon scattering. 
We have not found any previous mention of such effect in the phonon related literature.

\begin{figure}[t]
 \centering
  \includegraphics[width=8.3cm]{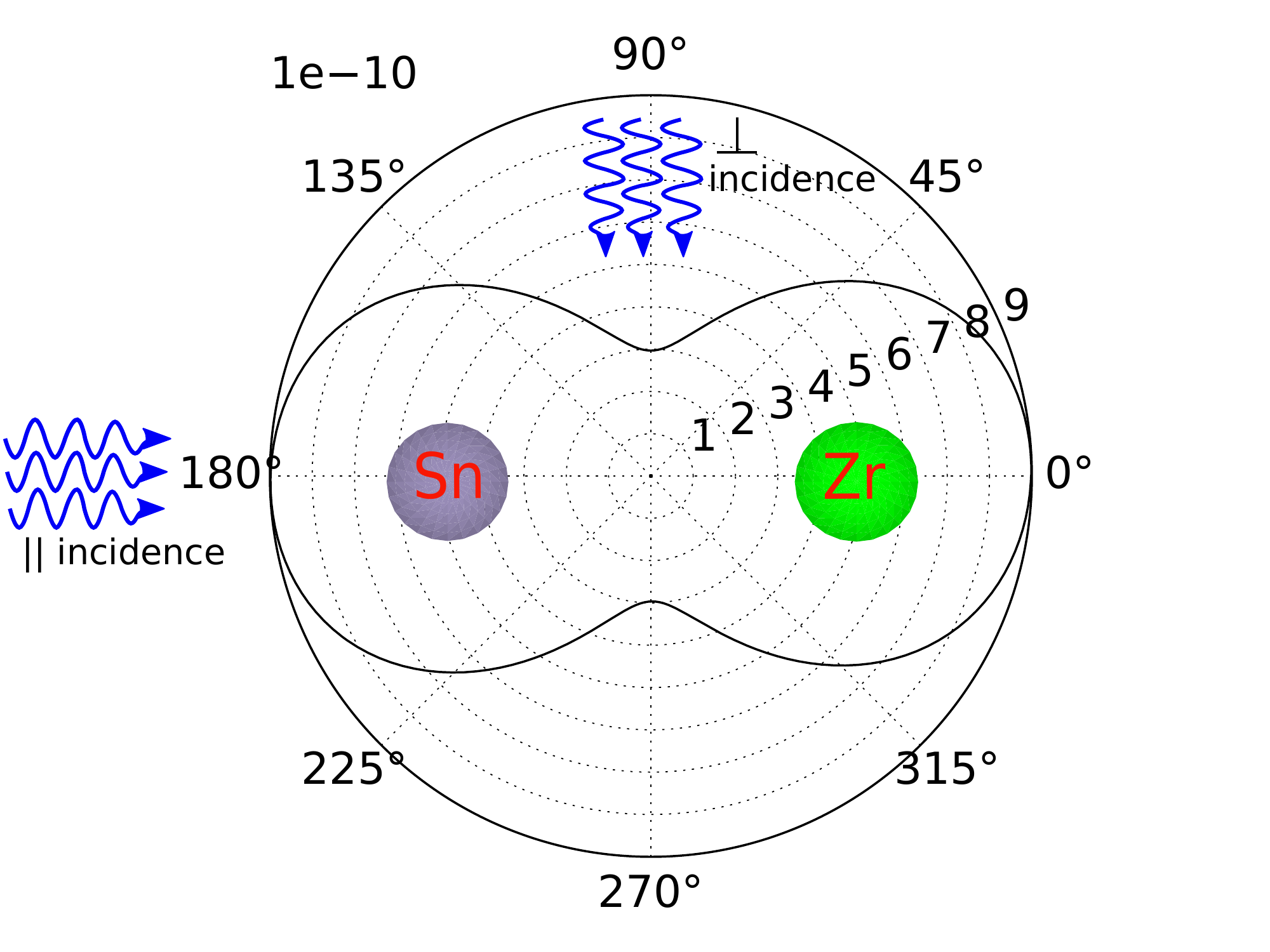}
  \caption{Scattering of LA phonons by a Sn/Zr antisite defect in ZrNiSn as a function of the angle
    between the wave vector and and the antisite bond. Unlike the case of a point defect, here scattering
    is markedly anisotropic. The strongest scattering is observed for an incidence parallel to the
    antisite bond, with the lowest scattering detected at $90^\circ$ from that direction.}
  \label{fig:pol_scat}
\end{figure}

\subsection{$\omega^6$ trend of antisite scattering.~~}
To gain some insight into the $\omega^6$ dependence of the antisite scattering rates, 
one can use the Born approximation, \emph{i.e.}, replace $(\mathbf{I -  
\hat{V}g^{+})^{-1}\hat{V}} \simeq \mathbf{\hat{V}}$ in Eq. \eqref{eq:tau_def}. 
This yields:

\begin{equation}
  1/\tau^{\text{def}}_{j\bq} \propto \left\vert \Braket{
      j'\bq' | \mathbf{\hat{V}} | j\bq } \right\vert^2 .
\label{eqn:fermi}
\end{equation}

For a long-wavelength acoustic phonon, the relevant projections of the wave functions on the two sites
forming the dipole, amount to a phase factor. Hence, the matrix element in Eq. \eqref{eqn:fermi} can be rewritten as:

\begin{equation}
  \Braket{j'\bq' | \mathbf{\hat{V}} | j\bq } = \Delta m_a  \omega^2 \exp(i\bq\cdot\bx) +  \Delta m_b \omega^2 \exp(-i\bq\cdot\bx),
\label{eqn:Vdipole}
\end{equation}

\noindent where $\bx$ is the vector connecting the two atoms involved, and $\Delta m_a$ and
$\Delta m_b$ are the mass differences introduced by the defect at both sites. For an antisite,
$\Delta m_a = (m_a - m_b) = -\Delta m_b$ where $m_a$ and $m_b$ are the original masses of the two atoms
that are exchanged. We can thus write Eq.~\eqref{eqn:Vdipole} as
\begin{equation}
  \Braket{j'\bq' | \mathbf{\hat{V}} | j\bq } = 2i \Delta m \omega^2 \sin(\bq\cdot\bx)
\end{equation}
In the limit of long wavelengths \emph{i.e.} very small $\left\vert\bq\right\vert$, we have
$\bq\cdot\bx \ll 1 \Rightarrow \sin(\bq\cdot\bx) \sim \bq\cdot\bx$ which leads to $\omega \propto  \left\vert\bq\right\vert$. 
Thus at low frequencies,
\begin{equation}
  \Braket{j'\bq' | \mathbf{\hat{V}} | j\bq } \propto \omega^3 \Delta m
\end{equation}
and we have
\begin{equation}
 1/ \tau^{\text{dipole def}} \propto \omega^6 .
\end{equation}
which is seen in the results for Sn/Zr and Ni/vacancy antisites in Fig.~\ref{fig:scat_rates}.
Along with the frequency dependence, another difference between the dipole and point defect is
the anisotropy of $1/ \tau^{\text{dipole~def}}$. Fig.~\ref{fig:pol_scat} shows the anisotropic
scattering rate calculated using a circular grid around the Sn/Zr dipole. The angular minimum
of the calculated scattering rates is not exactly zero as predicted by the analytical approximation, showing the
importance of full-blown Green's function calculations in obtaining accurate pictures of the scattering
process. Note that for obtaining the total scattering rate shown in Fig.~\ref{fig:scat_rates}, we have
averaged over all possible orientations of antisite defects introduced at random in the crystal.

\begin{figure}[t]
 \centering
  \includegraphics[width=8.3cm]{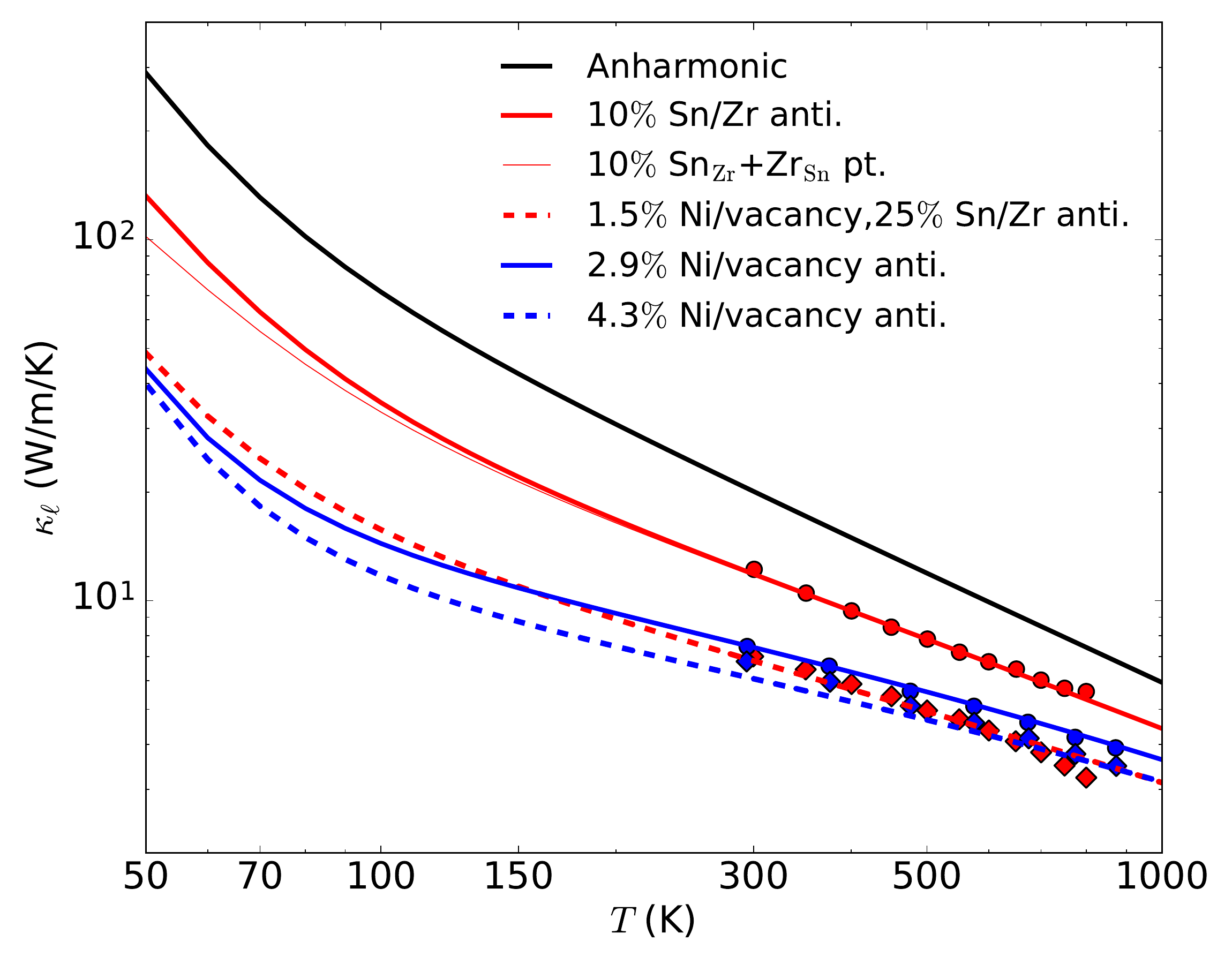}
  \caption{Calculated thermal conductivity of ZrNiSn with anharmonic phonon scattering only, with
    Ni/vacancy antisites and Sn/Zr antisites (bold lines) or with the independent
    antisite model (thin line). Experimental results from Ref.\citenum{Qiu_APL2010} (red markers) 
    and Ref.\citenum{Xie_CEC2012} (blue markers) are also shown for comparison. The '$\circ$' represent 
    $\kappa_{\ell}$ data for annealed and '$\diamond$' for unannealed samples.}
  \label{fig:kappa_l}
\end{figure}

The other important result we find from Fig.~\ref{fig:scat_rates} is that the scattering strength of the
Ni/vacancy antisites at low frequencies is $\sim 2$ orders of magnitude higher than that of Sn/Zr antisites in equal
concentration, which answers our first two questions. The higher scattering rate for Ni/vacancy
antisites can be attributed to the strong and long-ranged variations in the IFCs as compared to the
localised mass perturbation for the case of Sn/Zr antisites. Thus Ni/vacancy antisites will lead to
bigger thermal conductivity reduction as compared to the Sn/Zr antisites in ZrNiSn.

\subsection{Thermal conductivity of antisite enriched ZrNiSn.~~}
To answer our third question, we calculate the $\kappa_{\ell}$ for ZrNiSn with Ni/vacancy and Sn/Zr
antisites as shown in Fig.~\ref{fig:kappa_l}.  The experimentally determined concentrations for both
kinds of antisites are used for comparing our results to the experiments.\cite{Xie_CEC2012, Qiu_APL2010,
Aliev_ZPB1989, Aliev_ZPB1990} A very good agreement of the calculated $\kappa_{\ell}$ is found 
with the experiments by Xie \emph{et al.} for both annealed and unannealed samples.\cite{Xie_CEC2012} These 
results also show the dominant behaviour of Ni/vacancy antisites which, for annealed sample with $2.9\%$ concentration, 
brings down $\kappa_{\ell}$ by a factor of $\sim3$ as compared to the pure structure case, Fig.~\ref{fig:kappa_l}. 

To reproduce the results by Qiu \emph{et al.}, who have not determined the antisite concentration in
their samples, we have referred to the work of Aliev \emph{et al.} based on similar sample preparation 
technique. Surprisingly, Aliev \emph{et al.} have reported the presence of both Ni/vacancy and Sn/Zr antisites in 
their unannealed sample prepared with arc-melting method which contradicts the claim made by 
Qiu \emph{et al}.\cite{Aliev_ZPB1990, Qiu_APL2010} 
This is also confirmed by our calculations where to get $\kappa_{\ell}$ near to the experimental results for 
the unannealed samples by Qiu \emph{et al.}, 
we require an unrealistically high $\sim50\%$ Sn/Zr antisite concentration. 
Furthermore, Aliev \emph{et al.} found no Ni/vacancy but a 
substantial amount ($\sim10\%$) of Sn/Zr antisites even after annealing.\cite{Aliev_ZPB1989} 
This justifies the $\kappa_{\ell}$ results for annealed samples by Qiu \emph{et al.}, Fig.~\ref{fig:kappa_l},
which show strong deviatation from $1/T$ trend of $\kappa_{\ell}-T$ curve, that would be otherwise expected for nearly pure samples. 
Our calculated $\kappa_{\ell}$, including the pair antisite scattering contributions from both Ni/vacancy and 
Sn/Zr antisites for unannealed and only Sn/Zr antisites for annealed samples, are in good agreement 
with the results by Qiu \emph{et al.}, Fig.~\ref{fig:kappa_l}. The antisite concentrations used in our calculations are 
in similar range as suggested in Refs.~\citenum{Aliev_ZPB1989, Aliev_ZPB1990}. 
We also calculate $\kappa_{\ell}$ for the
independent antisite model and the differences from the pair antisite results are noticeable mainly at low
temperatures. This comes from the different low-frequency scattering trends for independent and pair
antisites, shown in Fig.~\ref{fig:scat_rates}, which dominate at low temperature.  We have not included
the boundary scattering effect in this study, which would reduce the low-temperature
$\kappa_{\ell}$. However, in the high-temperature range the antisite scattering, especially from
Ni/vacancy antisites, is found to be the dominant scattering mechanism in ZrNiSn. 
Thus, getting rid of Ni/vacancy antisites completely from the ZrNiSn can substantially increase $\kappa_{\ell}$, 
and vise-versa. Furthermore, purifying the annealed arc-melted samples with Sn/Zr antisites 
can also increase $\kappa_{\ell}$ by $\sim2$ times, Fig.~\ref{fig:kappa_l}. 

\section{\label{sec:conc}Conclusions}
We have developed an \emph{ab-initio} theoretical approach to determine antisite scattering strength in ZrNiSn half-Heusler 
compound and calculated the $\kappa_{\ell}$ within BTE-RTA framework. The $\kappa_{\ell}$ results for 
different samples of antisite-enriched ZrNiSn are in excellent agreement with previous experiments. Our results 
show that Ni/vacancy antisites, which are found to induce strong and long-ranged variations in the atomic environment,  
are the dominant scatterers in ZrNiSn as compared to, previously considered to be stronger,  
Sn/Zr antisites which effectively result in localized mass 
perturbations. Additionally, an $\omega^6$ dependence of phonon-antisite scattering rates is analytically 
derived and reproduced in our calculations. These results can be of technological interest for tuning the 
thermal conductivity of ZrNiSn, according to the application, by changing the antisite concentrations.
Furthermore, good agreement of our results with experiments suggest that the thermal conductivity measurements 
and \emph{ab-initio} calculations can be advantageously used together to determine sample defect concentrations 
in other materials as well.\cite{Oh_APL2011} 

\section{Acknowledgement}
The authors acknowledge the support from the Air Force Office of Scientific Research, USAF under award 
no. FA9550615-1-0187 DEF, the European Union's Horizon 2020 Research and Innovation Programme 
[grant number 645776 (ALMA)] , and ANR through project Carnot SIEVE.

\begin{mcitethebibliography}{34}
\providecommand*{\natexlab}[1]{#1}
\providecommand*{\mciteSetBstSublistMode}[1]{}
\providecommand*{\mciteSetBstMaxWidthForm}[2]{}
\providecommand*{\mciteBstWouldAddEndPuncttrue}
  {\def\EndOfBibitem{\unskip.}}
\providecommand*{\mciteBstWouldAddEndPunctfalse}
  {\let\EndOfBibitem\relax}
\providecommand*{\mciteSetBstMidEndSepPunct}[3]{}
\providecommand*{\mciteSetBstSublistLabelBeginEnd}[3]{}
\providecommand*{\EndOfBibitem}{}
\mciteSetBstSublistMode{f}
\mciteSetBstMaxWidthForm{subitem}
{(\emph{\alph{mcitesubitemcount}})}
\mciteSetBstSublistLabelBeginEnd{\mcitemaxwidthsubitemform\space}
{\relax}{\relax}

\bibitem[Freysoldt \emph{et~al.}(2014)Freysoldt, Grabowski, Hickel, Neugebauer,
  Kresse, Janotti, and Van~de Walle]{Freysoldt_2014}
C.~Freysoldt, B.~Grabowski, T.~Hickel, J.~Neugebauer, G.~Kresse, A.~Janotti and
  C.~G. Van~de Walle, \emph{Rev. Mod. Phys.}, 2014, \textbf{86}, 253--305\relax
\mciteBstWouldAddEndPuncttrue
\mciteSetBstMidEndSepPunct{\mcitedefaultmidpunct}
{\mcitedefaultendpunct}{\mcitedefaultseppunct}\relax
\EndOfBibitem
\bibitem[Hu and Tao(2015)]{Hu_JMCA2015}
B.~Hu and G.~Tao, \emph{J. Mater. Chem. A}, 2015, \textbf{3},
  20399--20407\relax
\mciteBstWouldAddEndPuncttrue
\mciteSetBstMidEndSepPunct{\mcitedefaultmidpunct}
{\mcitedefaultendpunct}{\mcitedefaultseppunct}\relax
\EndOfBibitem
\bibitem[Zandiatashbar \emph{et~al.}(2014)Zandiatashbar, Lee, An, Lee, Mathew,
  Terrones, Hayashi, Picu, Hone, and Koratkar]{Zandi_NComm2014}
A.~Zandiatashbar, G.-H. Lee, S.~J. An, S.~Lee, N.~Mathew, M.~Terrones,
  T.~Hayashi, C.~R. Picu, J.~Hone and N.~Koratkar, \emph{Nature
  Communications}, 2014, \textbf{5}, 3186\relax
\mciteBstWouldAddEndPuncttrue
\mciteSetBstMidEndSepPunct{\mcitedefaultmidpunct}
{\mcitedefaultendpunct}{\mcitedefaultseppunct}\relax
\EndOfBibitem
\bibitem[Doak \emph{et~al.}(2015)Doak, Michel, and Wolverton]{Doak_JMCC2015}
J.~W. Doak, K.~J. Michel and C.~Wolverton, \emph{J. Mater. Chem. C}, 2015,
  \textbf{3}, 10630--10649\relax
\mciteBstWouldAddEndPuncttrue
\mciteSetBstMidEndSepPunct{\mcitedefaultmidpunct}
{\mcitedefaultendpunct}{\mcitedefaultseppunct}\relax
\EndOfBibitem
\bibitem[Xu \emph{et~al.}(2013)Xu, Wei, Xu, Fan, and Zheng]{Xu_JMCA2013}
L.~Xu, N.~Wei, X.~Xu, Z.~Fan and Y.~Zheng, \emph{J. Mater. Chem. A}, 2013,
  \textbf{1}, 2002--2010\relax
\mciteBstWouldAddEndPuncttrue
\mciteSetBstMidEndSepPunct{\mcitedefaultmidpunct}
{\mcitedefaultendpunct}{\mcitedefaultseppunct}\relax
\EndOfBibitem
\bibitem[Cao \emph{et~al.}(2015)Cao, Cheng, Bi, Zhao, Yuan, Liu, Li, Wang, and
  Che]{Cao_JMCA2015}
Q.~Cao, Y.-F. Cheng, H.~Bi, X.~Zhao, K.~Yuan, Q.~Liu, Q.~Li, M.~Wang and
  R.~Che, \emph{J. Mater. Chem. A}, 2015, \textbf{3}, 20051--20055\relax
\mciteBstWouldAddEndPuncttrue
\mciteSetBstMidEndSepPunct{\mcitedefaultmidpunct}
{\mcitedefaultendpunct}{\mcitedefaultseppunct}\relax
\EndOfBibitem
\bibitem[Van~de Walle and Neugebauer(2004)]{Walle_JAP2004}
C.~G. Van~de Walle and J.~Neugebauer, \emph{Journal of Applied Physics}, 2004,
  \textbf{95}, 3851--3879\relax
\mciteBstWouldAddEndPuncttrue
\mciteSetBstMidEndSepPunct{\mcitedefaultmidpunct}
{\mcitedefaultendpunct}{\mcitedefaultseppunct}\relax
\EndOfBibitem
\bibitem[Liu \emph{et~al.}(2014)Liu, Lv, Zhang, Liu, Zhu, Zong, and
  Zhu]{Liu_JMCA2015}
D.~Liu, Y.~Lv, M.~Zhang, Y.~Liu, Y.~Zhu, R.~Zong and Y.~Zhu, \emph{J. Mater.
  Chem. A}, 2014, \textbf{2}, 15377--15388\relax
\mciteBstWouldAddEndPuncttrue
\mciteSetBstMidEndSepPunct{\mcitedefaultmidpunct}
{\mcitedefaultendpunct}{\mcitedefaultseppunct}\relax
\EndOfBibitem
\bibitem[Casper \emph{et~al.}(2012)Casper, Graf, Chadov, Balke, and
  Felser]{Casper_SST2012}
F.~Casper, T.~Graf, S.~Chadov, B.~Balke and C.~Felser, \emph{Semiconductor
  Science and Technology}, 2012, \textbf{27}, 063001\relax
\mciteBstWouldAddEndPuncttrue
\mciteSetBstMidEndSepPunct{\mcitedefaultmidpunct}
{\mcitedefaultendpunct}{\mcitedefaultseppunct}\relax
\EndOfBibitem
\bibitem[Chen and Ren(2013)]{Chen_MT2013}
S.~Chen and Z.~Ren, \emph{Materials Today}, 2013, \textbf{16}, 387 -- 395\relax
\mciteBstWouldAddEndPuncttrue
\mciteSetBstMidEndSepPunct{\mcitedefaultmidpunct}
{\mcitedefaultendpunct}{\mcitedefaultseppunct}\relax
\EndOfBibitem
\bibitem[Fecher \emph{et~al.}(2016)Fecher, Rausch, Balke, Weidenkaff, and
  Felser]{Fecher_PSSA2016}
G.~H. Fecher, E.~Rausch, B.~Balke, A.~Weidenkaff and C.~Felser, \emph{physica
  status solidi (a)}, 2016, \textbf{213}, 716--731\relax
\mciteBstWouldAddEndPuncttrue
\mciteSetBstMidEndSepPunct{\mcitedefaultmidpunct}
{\mcitedefaultendpunct}{\mcitedefaultseppunct}\relax
\EndOfBibitem
\bibitem[Sakurada and Shutoh(2005)]{Sakurada_APL2005}
S.~Sakurada and N.~Shutoh, \emph{Applied Physics Letters}, 2005, \textbf{86},
  082105\relax
\mciteBstWouldAddEndPuncttrue
\mciteSetBstMidEndSepPunct{\mcitedefaultmidpunct}
{\mcitedefaultendpunct}{\mcitedefaultseppunct}\relax
\EndOfBibitem
\bibitem[Shen \emph{et~al.}(2001)Shen, Chen, Goto, Hirai, Yang, Meisner, and
  Uher]{Shen_APL2001}
Q.~Shen, L.~Chen, T.~Goto, T.~Hirai, J.~Yang, G.~P. Meisner and C.~Uher,
  \emph{Applied Physics Letters}, 2001, \textbf{79}, 4165--4167\relax
\mciteBstWouldAddEndPuncttrue
\mciteSetBstMidEndSepPunct{\mcitedefaultmidpunct}
{\mcitedefaultendpunct}{\mcitedefaultseppunct}\relax
\EndOfBibitem
\bibitem[Ouardi \emph{et~al.}(2010)Ouardi, Fecher, Balke, Schwall, Kozina,
  Stryganyuk, Felser, Ikenaga, Yamashita, Ueda, and Kobayashi]{Ouardi_APL2010}
S.~Ouardi, G.~H. Fecher, B.~Balke, M.~Schwall, X.~Kozina, G.~Stryganyuk,
  C.~Felser, E.~Ikenaga, Y.~Yamashita, S.~Ueda and K.~Kobayashi, \emph{Applied
  Physics Letters}, 2010, \textbf{97}, 252113\relax
\mciteBstWouldAddEndPuncttrue
\mciteSetBstMidEndSepPunct{\mcitedefaultmidpunct}
{\mcitedefaultendpunct}{\mcitedefaultseppunct}\relax
\EndOfBibitem
\bibitem[Krez \emph{et~al.}(2014)Krez, Schmitt, Jeffrey~Snyder, Felser, Hermes,
  and Schwind]{Krez_JMCA2014}
J.~Krez, J.~Schmitt, G.~Jeffrey~Snyder, C.~Felser, W.~Hermes and M.~Schwind,
  \emph{J. Mater. Chem. A}, 2014, \textbf{2}, 13513--13518\relax
\mciteBstWouldAddEndPuncttrue
\mciteSetBstMidEndSepPunct{\mcitedefaultmidpunct}
{\mcitedefaultendpunct}{\mcitedefaultseppunct}\relax
\EndOfBibitem
\bibitem[Sootsman \emph{et~al.}(2009)Sootsman, Chung, and
  Kanatzidis]{Sootsman_AC2009}
J.~Sootsman, D.~Chung and M.~Kanatzidis, \emph{Angewandte Chemie International
  Edition}, 2009, \textbf{48}, 8616--8639\relax
\mciteBstWouldAddEndPuncttrue
\mciteSetBstMidEndSepPunct{\mcitedefaultmidpunct}
{\mcitedefaultendpunct}{\mcitedefaultseppunct}\relax
\EndOfBibitem
\bibitem[Yu \emph{et~al.}(2009)Yu, Zhu, Shi, Zhang, Zhao, and He]{Yu_AM2009}
C.~Yu, T.-J. Zhu, R.-Z. Shi, Y.~Zhang, X.-B. Zhao and J.~He, \emph{Acta
  Materialia}, 2009, \textbf{57}, 2757 -- 2764\relax
\mciteBstWouldAddEndPuncttrue
\mciteSetBstMidEndSepPunct{\mcitedefaultmidpunct}
{\mcitedefaultendpunct}{\mcitedefaultseppunct}\relax
\EndOfBibitem
\bibitem[Qiu \emph{et~al.}(2010)Qiu, Yang, Huang, Chen, and Chen]{Qiu_APL2010}
P.~Qiu, J.~Yang, X.~Huang, X.~Chen and L.~Chen, \emph{Applied Physics Letters},
  2010, \textbf{96}, 152105\relax
\mciteBstWouldAddEndPuncttrue
\mciteSetBstMidEndSepPunct{\mcitedefaultmidpunct}
{\mcitedefaultendpunct}{\mcitedefaultseppunct}\relax
\EndOfBibitem
\bibitem[Xie \emph{et~al.}(2012)Xie, Mi, Hu, Lock, Chirstensen, Fu, Iversen,
  Zhao, and Zhu]{Xie_CEC2012}
H.-H. Xie, J.-L. Mi, L.-P. Hu, N.~Lock, M.~Chirstensen, C.-G. Fu, B.~B.
  Iversen, X.-B. Zhao and T.-J. Zhu, \emph{CrystEngComm}, 2012, \textbf{14},
  4467--4471\relax
\mciteBstWouldAddEndPuncttrue
\mciteSetBstMidEndSepPunct{\mcitedefaultmidpunct}
{\mcitedefaultendpunct}{\mcitedefaultseppunct}\relax
\EndOfBibitem
\bibitem[Aliev \emph{et~al.}(1989)Aliev, Brandt, Moshchalkov, Kozyrkov,
  Skolozdra, and Belogorokhov]{Aliev_ZPB1989}
F.~G. Aliev, N.~B. Brandt, V.~V. Moshchalkov, V.~V. Kozyrkov, R.~V. Skolozdra
  and A.~I. Belogorokhov, \emph{Zeitschrift f{\"u}r Physik B Condensed Matter},
  1989, \textbf{75}, 167--171\relax
\mciteBstWouldAddEndPuncttrue
\mciteSetBstMidEndSepPunct{\mcitedefaultmidpunct}
{\mcitedefaultendpunct}{\mcitedefaultseppunct}\relax
\EndOfBibitem
\bibitem[Li \emph{et~al.}(2012)Li, Lindsay, Broido, Stewart, and
  Mingo]{Li_PRB2012}
W.~Li, L.~Lindsay, D.~A. Broido, D.~A. Stewart and N.~Mingo, \emph{Phys. Rev.
  B}, 2012, \textbf{86}, 174307\relax
\mciteBstWouldAddEndPuncttrue
\mciteSetBstMidEndSepPunct{\mcitedefaultmidpunct}
{\mcitedefaultendpunct}{\mcitedefaultseppunct}\relax
\EndOfBibitem
\bibitem[Katre and Madsen(2016)]{Katre_PRB2016}
A.~Katre and G.~K.~H. Madsen, \emph{Phys. Rev. B}, 2016, \textbf{93},
  155203\relax
\mciteBstWouldAddEndPuncttrue
\mciteSetBstMidEndSepPunct{\mcitedefaultmidpunct}
{\mcitedefaultendpunct}{\mcitedefaultseppunct}\relax
\EndOfBibitem
\bibitem[Mingo \emph{et~al.}(2010)Mingo, Esfarjani, Broido, and
  Stewart]{Mingo_PRB2010}
N.~Mingo, K.~Esfarjani, D.~A. Broido and D.~A. Stewart, \emph{Phys. Rev. B},
  2010, \textbf{81}, 045408\relax
\mciteBstWouldAddEndPuncttrue
\mciteSetBstMidEndSepPunct{\mcitedefaultmidpunct}
{\mcitedefaultendpunct}{\mcitedefaultseppunct}\relax
\EndOfBibitem
\bibitem[Katcho \emph{et~al.}(2014)Katcho, Carrete, Li, and
  Mingo]{Katcho_PRB2014}
N.~A. Katcho, J.~Carrete, W.~Li and N.~Mingo, \emph{Phys. Rev. B}, 2014,
  \textbf{90}, 094117\relax
\mciteBstWouldAddEndPuncttrue
\mciteSetBstMidEndSepPunct{\mcitedefaultmidpunct}
{\mcitedefaultendpunct}{\mcitedefaultseppunct}\relax
\EndOfBibitem
\bibitem[Kundu \emph{et~al.}(2011)Kundu, Mingo, Broido, and
  Stewart]{Kundu_PRB2011}
A.~Kundu, N.~Mingo, D.~A. Broido and D.~A. Stewart, \emph{Phys. Rev. B}, 2011,
  \textbf{84}, 125426\relax
\mciteBstWouldAddEndPuncttrue
\mciteSetBstMidEndSepPunct{\mcitedefaultmidpunct}
{\mcitedefaultendpunct}{\mcitedefaultseppunct}\relax
\EndOfBibitem
\bibitem[Bl{\"o}chl(1994)]{paw}
P.~E. Bl{\"o}chl, \emph{Phys. Rev. B}, 1994, \textbf{50}, 17953--17979\relax
\mciteBstWouldAddEndPuncttrue
\mciteSetBstMidEndSepPunct{\mcitedefaultmidpunct}
{\mcitedefaultendpunct}{\mcitedefaultseppunct}\relax
\EndOfBibitem
\bibitem[Kresse and Joubert(1999)]{PAW_method}
G.~Kresse and D.~Joubert, \emph{Phys. Rev. B}, 1999, \textbf{59}, 1758\relax
\mciteBstWouldAddEndPuncttrue
\mciteSetBstMidEndSepPunct{\mcitedefaultmidpunct}
{\mcitedefaultendpunct}{\mcitedefaultseppunct}\relax
\EndOfBibitem
\bibitem[Togo \emph{et~al.}(2008)Togo, Oba, and Tanaka]{phonopy}
A.~Togo, F.~Oba and I.~Tanaka, \emph{Phys. Rev. B}, 2008, \textbf{78},
  134106\relax
\mciteBstWouldAddEndPuncttrue
\mciteSetBstMidEndSepPunct{\mcitedefaultmidpunct}
{\mcitedefaultendpunct}{\mcitedefaultseppunct}\relax
\EndOfBibitem
\bibitem[Li \emph{et~al.}(2014)Li, Carrete, Katcho, and Mingo]{Li_CPC2014}
W.~Li, J.~Carrete, N.~A. Katcho and N.~Mingo, \emph{Computer Physics
  Communications}, 2014, \textbf{185}, 1747 -- 1758\relax
\mciteBstWouldAddEndPuncttrue
\mciteSetBstMidEndSepPunct{\mcitedefaultmidpunct}
{\mcitedefaultendpunct}{\mcitedefaultseppunct}\relax
\EndOfBibitem
\bibitem[Wang \emph{et~al.}(2010)Wang, Wang, Wang, Mei, Shang, Chen, and
  Liu]{Wang_JPCM2010}
Y.~Wang, J.~J. Wang, W.~Y. Wang, Z.~G. Mei, S.~L. Shang, L.~Q. Chen and Z.~K.
  Liu, \emph{Journal of Physics: Condensed Matter}, 2010, \textbf{22},
  202201\relax
\mciteBstWouldAddEndPuncttrue
\mciteSetBstMidEndSepPunct{\mcitedefaultmidpunct}
{\mcitedefaultendpunct}{\mcitedefaultseppunct}\relax
\EndOfBibitem
\bibitem[Lambin and Vigneron(1984)]{tetrahedron}
P.~Lambin and J.~P. Vigneron, \emph{Phys. Rev. B}, 1984, \textbf{29},
  3430--3437\relax
\mciteBstWouldAddEndPuncttrue
\mciteSetBstMidEndSepPunct{\mcitedefaultmidpunct}
{\mcitedefaultendpunct}{\mcitedefaultseppunct}\relax
\EndOfBibitem
\bibitem[ALM()]{ALMA_BTE}
\url{http://www.almabte.eu/}\relax
\mciteBstWouldAddEndPuncttrue
\mciteSetBstMidEndSepPunct{\mcitedefaultmidpunct}
{\mcitedefaultendpunct}{\mcitedefaultseppunct}\relax
\EndOfBibitem
\bibitem[Aliev \emph{et~al.}(1990)Aliev, Kozyrkov, Moshchalkov, Scolozdra, and
  Durczewski]{Aliev_ZPB1990}
F.~G. Aliev, V.~V. Kozyrkov, V.~V. Moshchalkov, R.~V. Scolozdra and
  K.~Durczewski, \emph{Zeitschrift f{\"u}r Physik B Condensed Matter}, 1990,
  \textbf{80}, 353--357\relax
\mciteBstWouldAddEndPuncttrue
\mciteSetBstMidEndSepPunct{\mcitedefaultmidpunct}
{\mcitedefaultendpunct}{\mcitedefaultseppunct}\relax
\EndOfBibitem
\bibitem[Oh \emph{et~al.}(2011)Oh, Ravichandran, Liang, Siemons, Jalan, Brooks,
  Huijben, Schlom, Stemmer, Martin, Majumdar, Ramesh, and Cahill]{Oh_APL2011}
D.-W. Oh, J.~Ravichandran, C.-W. Liang, W.~Siemons, B.~Jalan, C.~M. Brooks,
  M.~Huijben, D.~G. Schlom, S.~Stemmer, L.~W. Martin, A.~Majumdar, R.~Ramesh
  and D.~G. Cahill, \emph{Applied Physics Letters}, 2011, \textbf{98},
  221904\relax
\mciteBstWouldAddEndPuncttrue
\mciteSetBstMidEndSepPunct{\mcitedefaultmidpunct}
{\mcitedefaultendpunct}{\mcitedefaultseppunct}\relax
\EndOfBibitem
\end{mcitethebibliography}

\providecommand*{\mcitethebibliography}{\thebibliography}
\csname @ifundefined\endcsname{endmcitethebibliography}
{\let\endmcitethebibliography\endthebibliography}{}

\end{document}